\documentclass[journal]{IEEEtran}
\usepackage[dvipdfm]{graphicx}    

\begin{document}
\title{Development of Particle-in-Cell Simulation in a Two Dimensional Trench Geometry}
\author{Tai-Lu Lin and Yasutaro Nishimura}
\thanks{The authors are with Institute of Space and Plasma Sciences,
National Cheng Kung University, Tainan 70101, Taiwan (e-mail: nishimura@pssc.ncku.edu.tw).}
\thanks{Manuscript received March 31, 2016.}
\maketitle

\begin{abstract}
A two dimensional electrostatic Particle-in-Cell simulation code is developed to investigate
anisotropy of ions in a trench geometry for plasma etching.
The numerical simulation results suggest that if the trench width 
is larger than Debye length scale, anisotropy can be lost due to potential development across the trench.
Furthermore, the effects of ion charge build up on the trench bottom is investigated,
which can degrade the anisotropy.
\end{abstract}


\section{Introduction}
\label{s1}

\IEEEPARstart{F}{or} semiconductor processing in industries, 
plasma etching has been one of the key technologies. 
We can categorize the material etching into two types: dry etching and wet etching \cite{lie05}. 
Wet etching employs chemical reactions by liquid to etch materials, which 
is a relatively low cost process. However, when the trench
size becomes small, it may cause undercut \cite{hit99} 
because etching direction becomes isotropic (meaning the etching is uniform in all directions). 
On the other hand, dry etching employs charged gases to etch materials. 
Ion bombardment resulting from
acceleration by sheath potential is the primary mechanism for dry etching. 
By the dry etching, anisotropic
etching can be achieved because the ion bombardment is primarily in one direction.

A capacitively coupled plasma (CCP) \cite{god90,tur95} is frequently employed for plasma etching.
A CCP can be seen as a plasma bounded by two plates.
One plate with an etching target is connected to the radio frequency (RF) bias and the other is grounded. 
Plasma discharge is sustained by stochastic heating in the sheath \cite{god92}.
When the plasma is generated, a sheath is developed near the material surface. 
The sheath accelerates ions and reflects electrons. 
Sheath formation is the key mechanism for anisotropic ion bombardment. 
On the other hand, within the trench, we do not want sheath to be developed in the
direction perpendicular to the trench.
Otherwise, the ions accelerated by the wall can erode sidewalls of the trench.
In this work, the plasma behavior in the trench
geometry is investigated by varying its size. 
We demonstrate that the trench width need to be 
smaller than a few Debye length. 

The industrial trend shows that the etching scale is becoming
smaller and smaller to reduce computational power \cite{smi16}, 
and improve memory size. 
To resolve sub-micron scale, one may need to rely on molecular dynamics approach which can be
numerically demanding, however. 
In this work, despite the scale is slightly larger (than the sub-micron scale),
we would like to establish a firm ground
of self-consistent charged particle dynamics in a trench geometry; 
whether ion anisotropy is kept or not.
Despite seemingly straightforward, Particle-in-Cell (PIC) simulation \cite{daw62,bir86} in a
realistic two-dimensional trench geometry is not conducted by many authors \cite{dai12}.
We consider the plasma discharge based on industrial setting between two plates
as in a capacitively coupled plasma and investigate the stable discharge for etching process in 
the trench geometry.

This paper is organized as follows.
In Section \ref{s2}, the basic computational model in 
the presence of kinetic ions and the electrons is described. 
In Section \ref{s3}, a series of two dimensional Particle-in-Cell simulation, 
that are in a square geometry,
in a periodic square geometry, and finally with the trench geometry, are discussed. 
We summarize this work in Section \ref{s4}.

\section{Computational model}
\label{s2}
In this section, our computational model of two dimensional electrostatic
PIC simulation is described. In the numerical simulation we incorporate both
the electron and an ion species. 
We extend our previous work in \cite{hua15} from one dimension to two dimension.

In this paper, we denote $n_0$ as the background plasma density and
$T_e$ ($T_i$) as the electron (ion) temperature. 
We denote $m_j$ and $q_j$ as the mass and the charge of the species
($j=e$ for electrons and $j=i$ for ions),
and $e = - q_e$ as the unit charge. 
The Debye length is given by
$\lambda_e = \left( \varepsilon_0 T_e / n_0 e^2 \right)^{1/2}$
and the plasma frequency is given by
$\omega_e = \left( n_0 e^2 / \varepsilon_0 m_e \right)^{1/2}$,
where $\varepsilon_0 $ is the vacuum permittivity. 
We employ the MKS unit in this paper.
As in \cite{hua15}, after normalizing length by Debye length,
time by the inverse of plasma frequency \cite{nic92},
and the electrostatic potential by $ T_e / e $,
the equations of motion are given by
\begin{equation}
\frac{d \bar{v_x}}{d \bar{t}} =  \alpha_j \bar{E_x} \left(  \bar{x} , \bar{y} \right)
\label{newton1}
\end{equation}
\begin{equation}
\frac{d \bar{v_y}}{d \bar{t}} =  \alpha_j \bar{E_y} \left(  \bar{x} , \bar{y} \right)
\label{newton2}
\end{equation}
\begin{equation}
\frac{d \bar{x}}{d \bar{t}} =  \bar{v}_x
\label{newton3}
\end{equation}
and
\begin{equation}
\frac{d \bar{y}}{d \bar{t}} =  \bar{v}_y
\label{newton4}
\end{equation}
where $\alpha_j = -1$ for $j=e$ and $\alpha_j = (m_e/m_i) (q_i/e)$ for $j=i$. 
Here, $(\bar{x}, \bar{y})$ and $(\bar{v}_x, \bar{v}_y)$ represent the coordinates 
of the configuration space and the velocity space, respectively. 
The normalized time variable is given by $\bar{t}$.
As a reminder the velocities are normalized by the electron thermal velocity
$v_e = \lambda_e \omega_e$.

On the other hand, normalized two dimensional Poisson equation is given by
(we normalize densities by $n_0$)
\begin{equation}
\left( \frac{\partial^2 } {  \partial {\bar{x}}^2 }  + 
\frac{\partial^2 } {  \partial {\bar{y}}^2 }  \right) 
\bar{\Phi}
=  \bar{n_e } - \frac{q_i}{e} \bar{n_i }   .
\label{poisson} 
\end{equation}
In (\ref{newton1})-(\ref{poisson}), all the quantities with ``bar'' 
denote normalized ones.
Here, $\bar{E}_x \left( \bar{x}, \bar{y} \right) = - \partial_{\bar{x}} \bar{\Phi} (\bar{x}, \bar{y})$ 
and $\bar{E_y} \left( \bar{x}, \bar{y} \right) = - \partial_{\bar{y}} \bar{\Phi} (\bar{x}, \bar{y})$
while $\bar{\Phi} (\bar{x}, \bar{y} )$ is the electrostatic potential.
For the numerical derivative, we take Euler differencing using two mesh points at the boundaries
(central differencing taken, otherwise).

In numerically integrating (\ref{newton1})-(\ref{newton4}), 
we employ the leapfrog method \cite{bir86,rut83}.
Because $\bar{x}$ and $\bar{y}$ dynamics are independent, we evolve in the order of
(\ref{newton1}) and (\ref{newton2}), (\ref{poisson}), (\ref{newton3}) and (\ref{newton4}).
In solving two dimensional Poisson equation (\ref{poisson}),
we employ a Gauss-Seidel method and a Jacobi method.
This is to assure correctness of the solution in the complicated trench geometries first, 
despite it may not be one of the fastest numerical schemes.
We employ a homogeneous Dirichlet boundary condition for the electrostatic potential
at one of the electrodes and a time dependent (an RF like) Dirichlet boundary condition
at the other electrode. 

\section{Two dimensional simulation results}
\label{s3}

In this section, we discuss our numerical simulation results by our newly developed two 
dimensional PIC code. To reach our final goal of simulating a realistic trench geometry, 
we build our models by verifying from a simple square geometry. We then impose periodic
condition in the direction parallel to the plates (which we refer to as
$\bar{y}$ direction), and finally incorporate a trench geometry.

Plasma parameters employed in the numerical simulation are:
plasma temperature $T_e=T_i$ (the temperature ratio $\tau = T_i / T_e =1$) 
and a mass ratio of $m_i/m_e=3672$.
For example, with $T_e= 10 eV$ and $n_0 = 2.27 \times 10^{10} cm^{-3}$, 
we obtain $\lambda_e = 1.55 \times 10^{-2} cm$ and $\omega_e = 8.52 \times 10^{9} s^{-1}$.
Then the industrial RF frequency typically used for capacitively coupled plasmas,
$13.56 MHz$, reads ${\bar{\omega}}_{RF}=0.01$ in our normalized units.
All the plasma dynamics in this work is regarded as collisionless.

\subsection{PIC simulation in a square geometry}
\label{s3a}

We first build a PIC model for a simple square geometry shown in Fig.~1.
In initiating the PIC simulation, we employ Gaussian for the velocity distribution 
for both $ {v}_x $ and $ {v}_y$ (see \cite{hua15} for methods to produce Gaussian):
\begin{equation}
f_{jv} ( {v}_x , {v}_y )  = \frac{n_0}{2 \pi v_j^2} \exp{\left[ -\frac{v_x^2 + v_y^2}{2 v_j^2} \right]}
\label{gaussian1} 
\end{equation}
where $j=e$ for electrons ($j=i$ for the ions)
and $v_i = v_e  \left( m_e \tau/ m_i \right)^{1/2} $ is the ion thermal velocity.

The initial particle positions in the configuration space are uniform
within $0 \le \bar{x} \le \bar{x}_{max}$ and $0 \le \bar{y} \le \bar{y}_{max}$. 
We distribute particles using a random number generator.
In Fig.~1, we take $\bar{x}_{max} = 60$ and $\bar{y}_{max} = 60$. 
The number of mesh points we take here are $n_x=60$ and $n_y=60$ 
(the mesh size is the Debye length).
We employ Dirichlet boundary condition on $\bar{x}= 0$, $\bar{x}=\bar{x}_{max}$, 
$\bar{y}=0$, and $\bar{y}=\bar{y}_{max}$  
(we set $\bar{\Phi} = 0$ assuming the four boundaries are grounded).  
Note that the electric field components along the surface of the plates are taken to be
zero (we assume conducting walls).
We take one hundred particles per two-dimensional-cell for both the ions and the electrons.
The time step is given by $ \Delta \bar{t} = 0.1$.

Sheath potential is developed \cite{hua15} as shown in Fig.~1(a) and electrons are
confined in two-dimensional square domain.
As shown in Fig.~1(b), high energy electrons are lost into the boundaries making the plasma 
positively charged at the steady state. To reach Fig.~1(a) and Fig.~1(b),
(1) to (5) are time advanced self-consistently up to $\bar{t} = 500$.
Ions are accelerated at the sheath due to large electric field toward the boundaries 
satisfying the Bohm-sheath criterion (as in \cite{hua15}) . 
In sustaining a steady state plasma discharge,
if an ion is lost at the boundaries, 
we provide the bulk plasma with a thermal ion
regarding it as an ionization of the ambient neutral gas.
In Fig.~1(b), four percent of total electrons are plotted.

Single electron orbit in a square geometry is shown in Fig.1(c).
As an unexpected by-product of the sheath dynamics study, 
we discover the orbit fill in the $\bar{x}-\bar{y}$ space uniformly.
This, as we observe, is similar to billiard balls reflecting at the elastic walls
and filling in the volume ergodically \cite{sto99}
(note that the electrons' {\it reflection} is given by deceleration and acceleration 
by the electric field near the boundaries).

\begin{figure}[ht]
\centering
\includegraphics[width=2.1in,angle=-90]{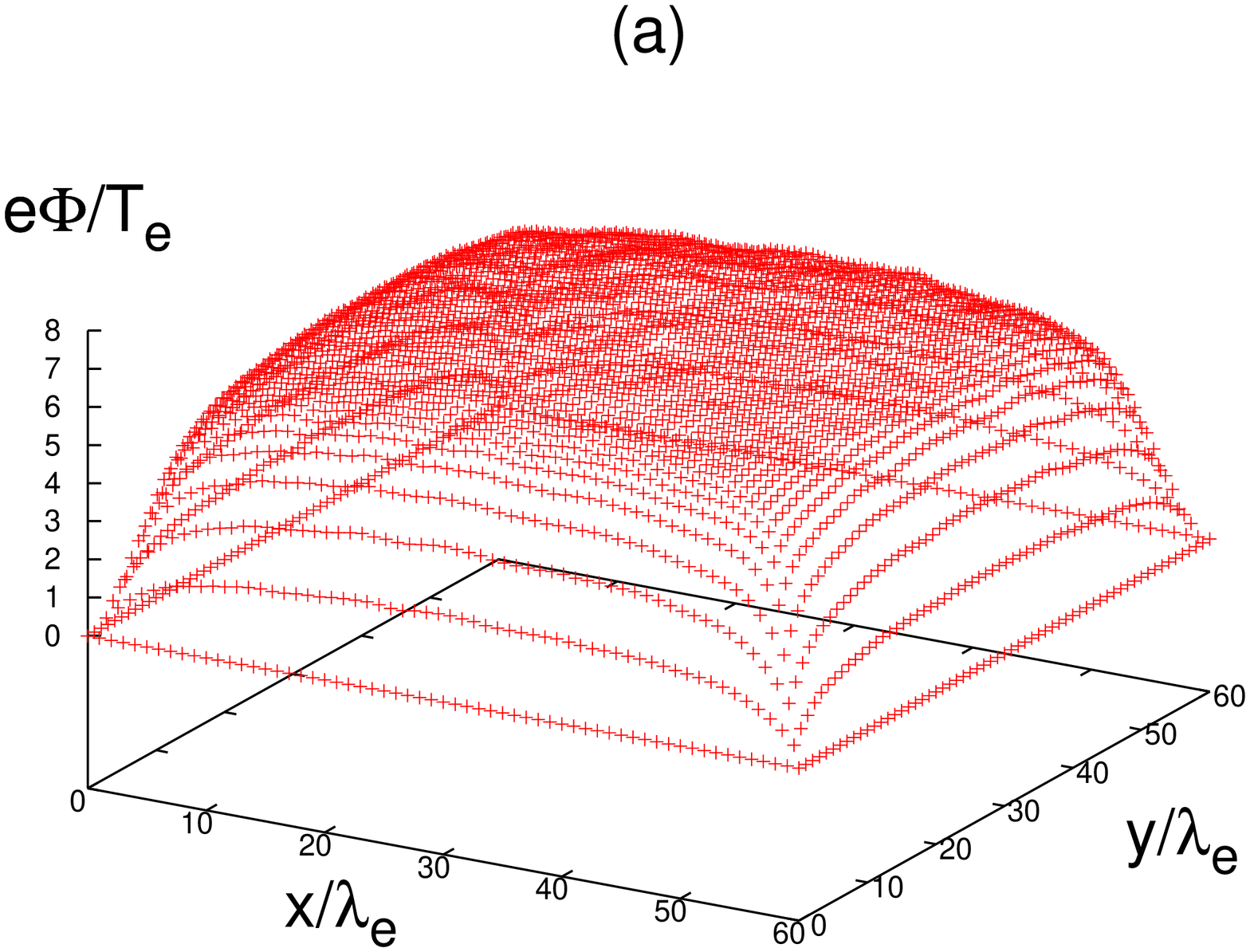}
\includegraphics[width=2.1in]{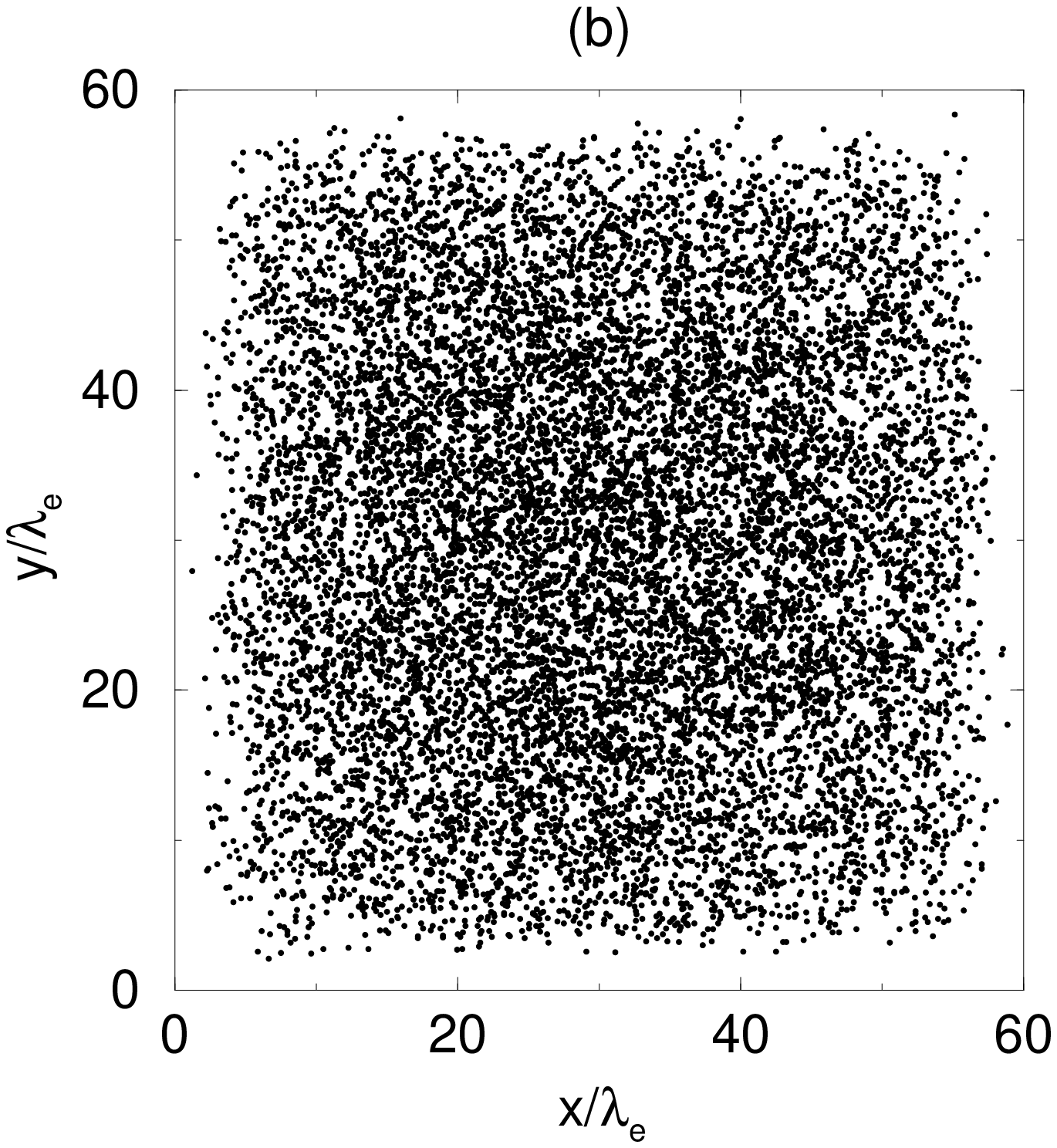}
\includegraphics[width=2.1in]{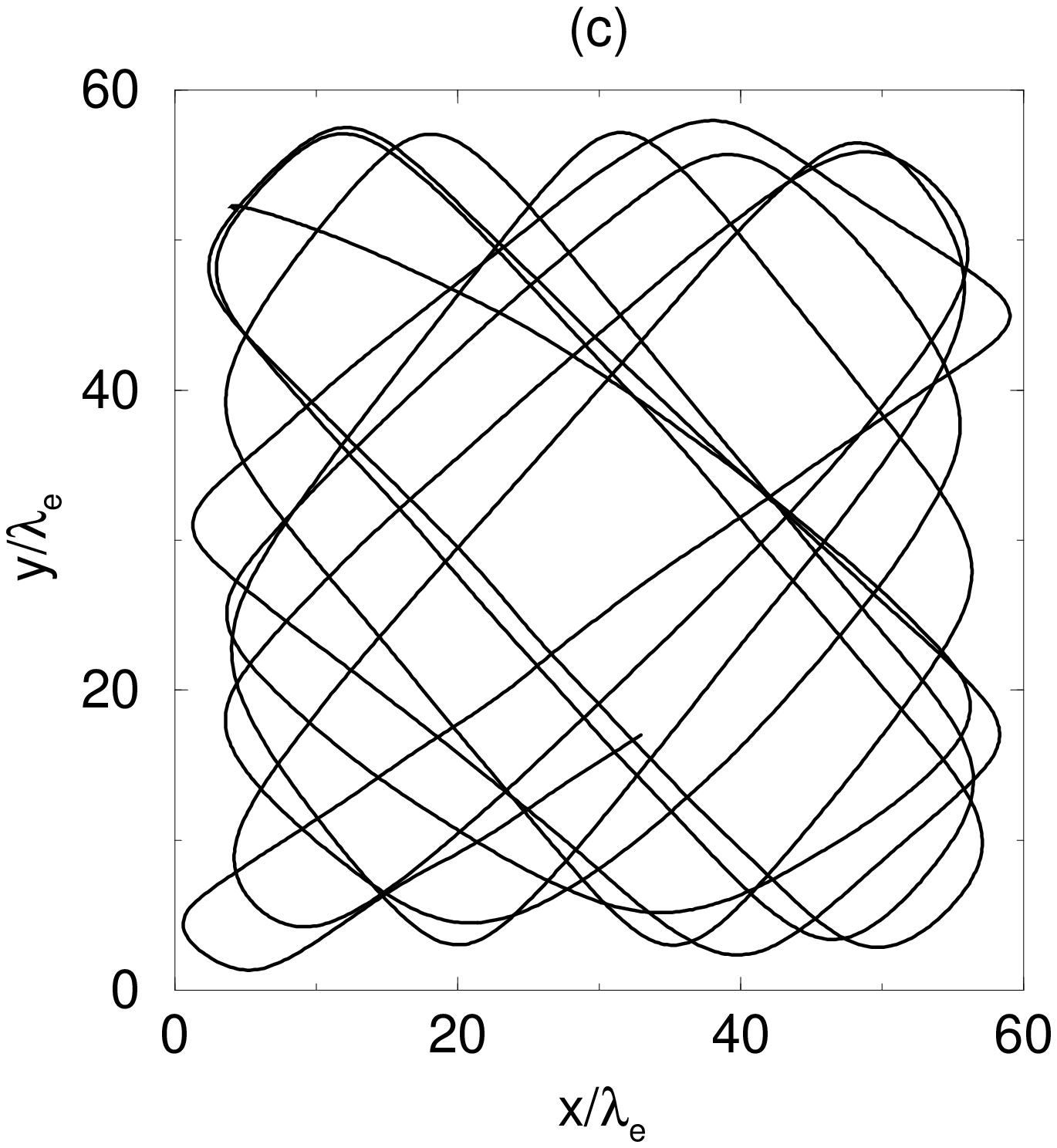}
\caption{ 
(a) A potential surface plot at the steady state at $\omega_e t = 500$.
(b) The electron distribution in the $\bar{x}-\bar{y}$ domain at $\omega_e t = 500$.
(c) An electron orbit in a square geometry which shows that an 
electron fills in the $\bar{x}-\bar{y}$ space ergodically.
The slight angle change during the reflection is due to 
the slow time variation of the electrostatic potential.
This specific orbit starts at $(\bar{x}, \bar{y}) = (32.9,17.0)$ 
and ends at $(\bar{x}, \bar{y})= (4.33,52.0)$.
}
\end{figure}

\subsection{PIC simulation with periodicity and an RF bias}
\label{s3b}
To further move on, we make the $\bar{y}$ direction periodic.
This is a second step toward the trench geometry.
This implies that the trench pattern repeats in the $\bar{y}$ direction
and the trench region we study are far away from the capacitor edges.
The parameters taken in Fig.2 is the same with Fig.1 except that $\bar{y}_{max}=30$
and the Poisson equation is solved by imposing periodicity at $\bar{y}=0$ and $\bar{y}=\bar{y}_{max}$. 
When the particles cross $\bar{y}=\bar{y}_{max}$,
they are designed to re-enter $\bar{y} =0$ and vice versa.
We still employ Dirichlet boundary conditions at $\bar{x}= 0$ ($\bar{\Phi} = 0$) 
and $\bar{x}=\bar{x}_{max} = 60$
[$ {\bar{\Phi}}  = {\bar{\Phi}}_0 \sin{ ( {\bar{\omega}}_{RF} \bar{t} ) } $ 
for an RF biased side. 
We take $\bar{\Phi}_0=5.0$ and ${\bar{\omega}}_{RF}=0.01$].  
Note that (since there are no obstacles so far, such as a trench)
the plasma dynamics should be equivalent to that of one dimensional system.

A surface plot of the electrostatic potential is shown in Fig. 2(a).
A comparison with one dimensional simulation results are given in Fig.~2(b) and Fig.~2(c).
Figure~2(b) compares $\bar{x}$ variation of $\bar{\Phi} $ 
($\bar{x}$ is the direction perpendicular to the plates, as a reminder) 
at fixed times $\bar{t} = 385$ and $\bar{t} = 435$,
and Fig.~2(c) compares the time dependence of potential values at $\bar{x}  = 50$. 
In Fig.~2(c), long time scale oscillation is by the bounce motion induced by
the RF boundary condition and small ripples originate from plasma oscillations.
The two data sets from one dimensional and two dimensional simulation compare favorably.
\begin{figure}[ht]
\centering
\includegraphics[width=2.1in,angle=-90]{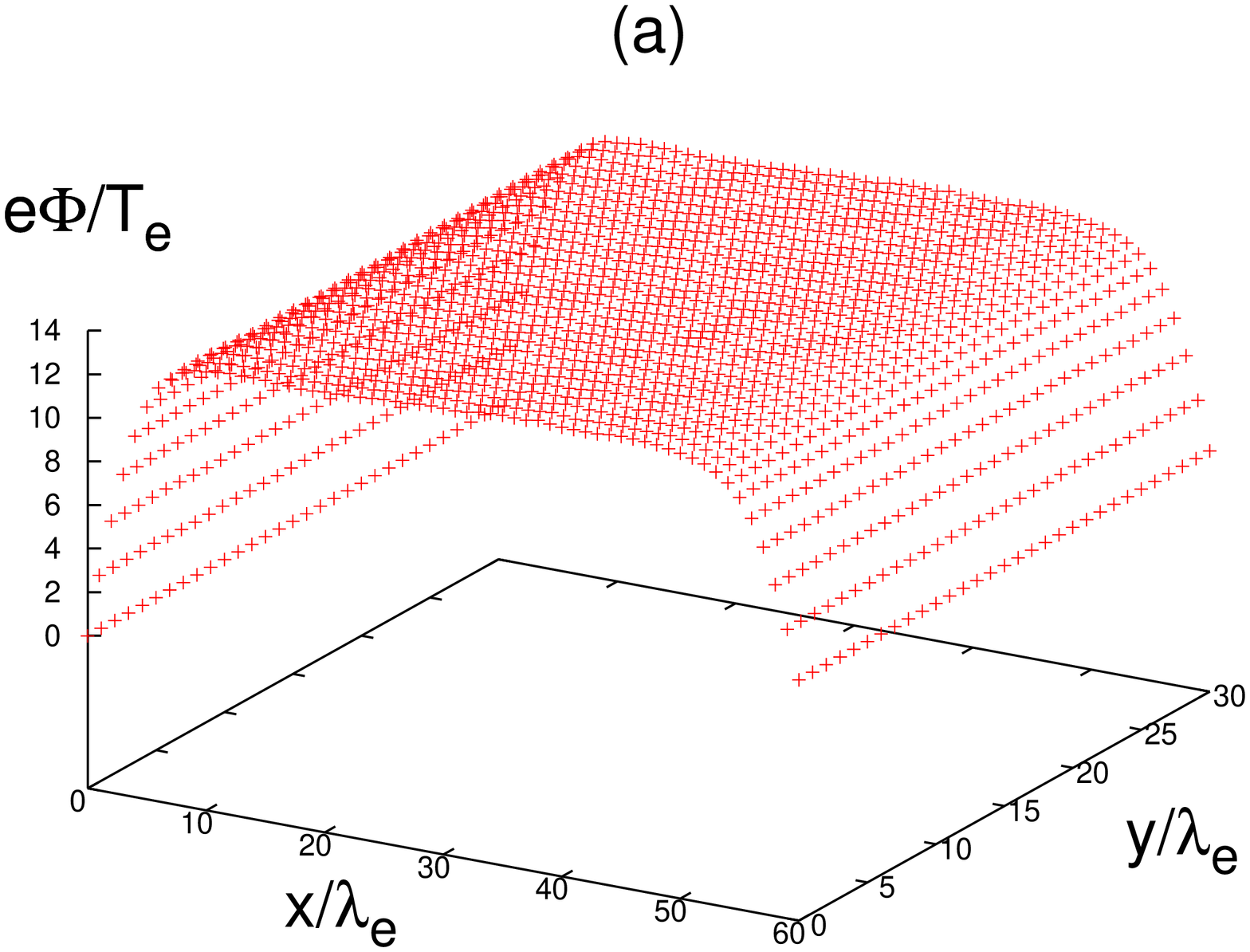}
\includegraphics[width=2.1in,angle=-00]{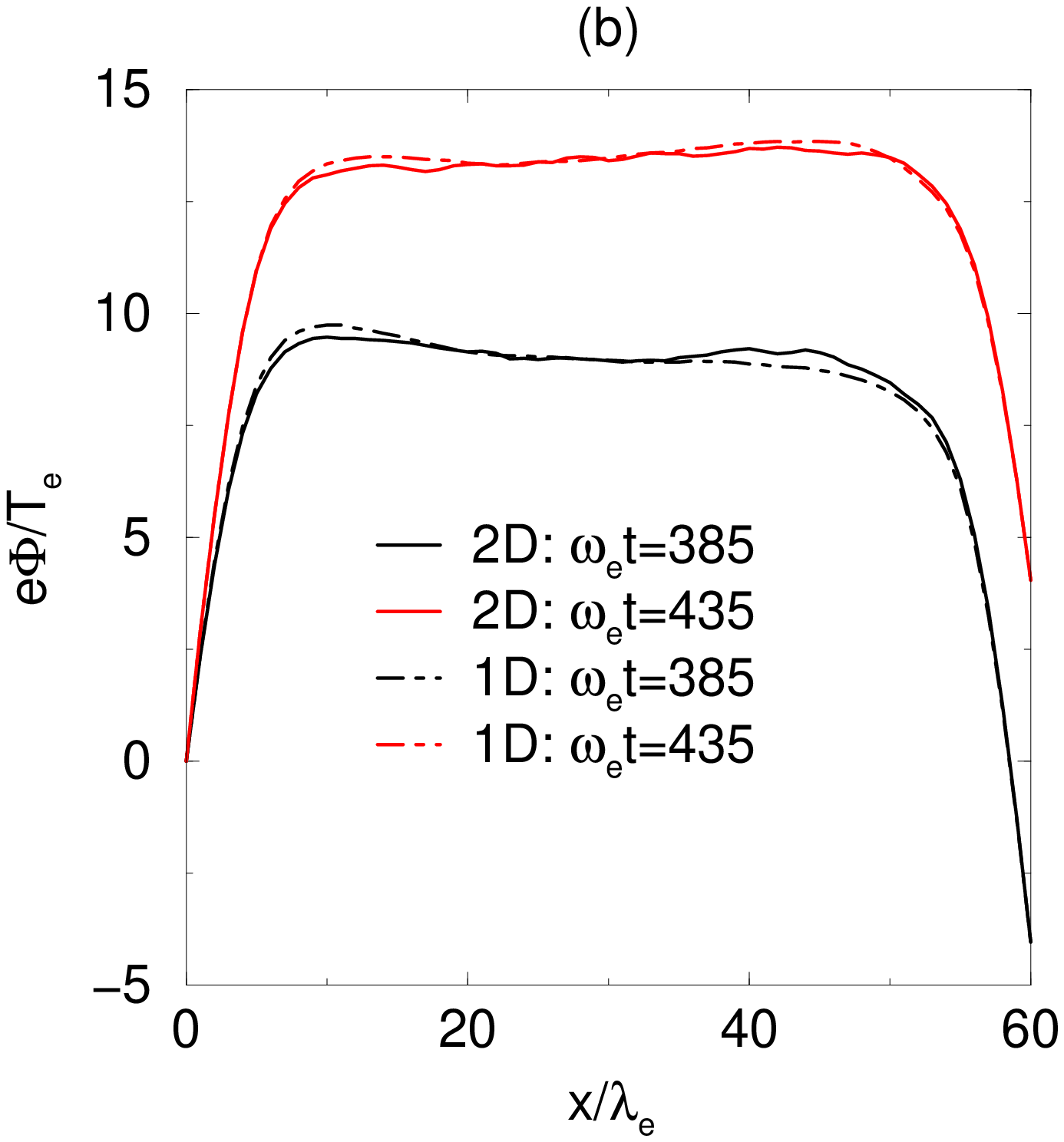}
\includegraphics[width=2.1in,angle=-00]{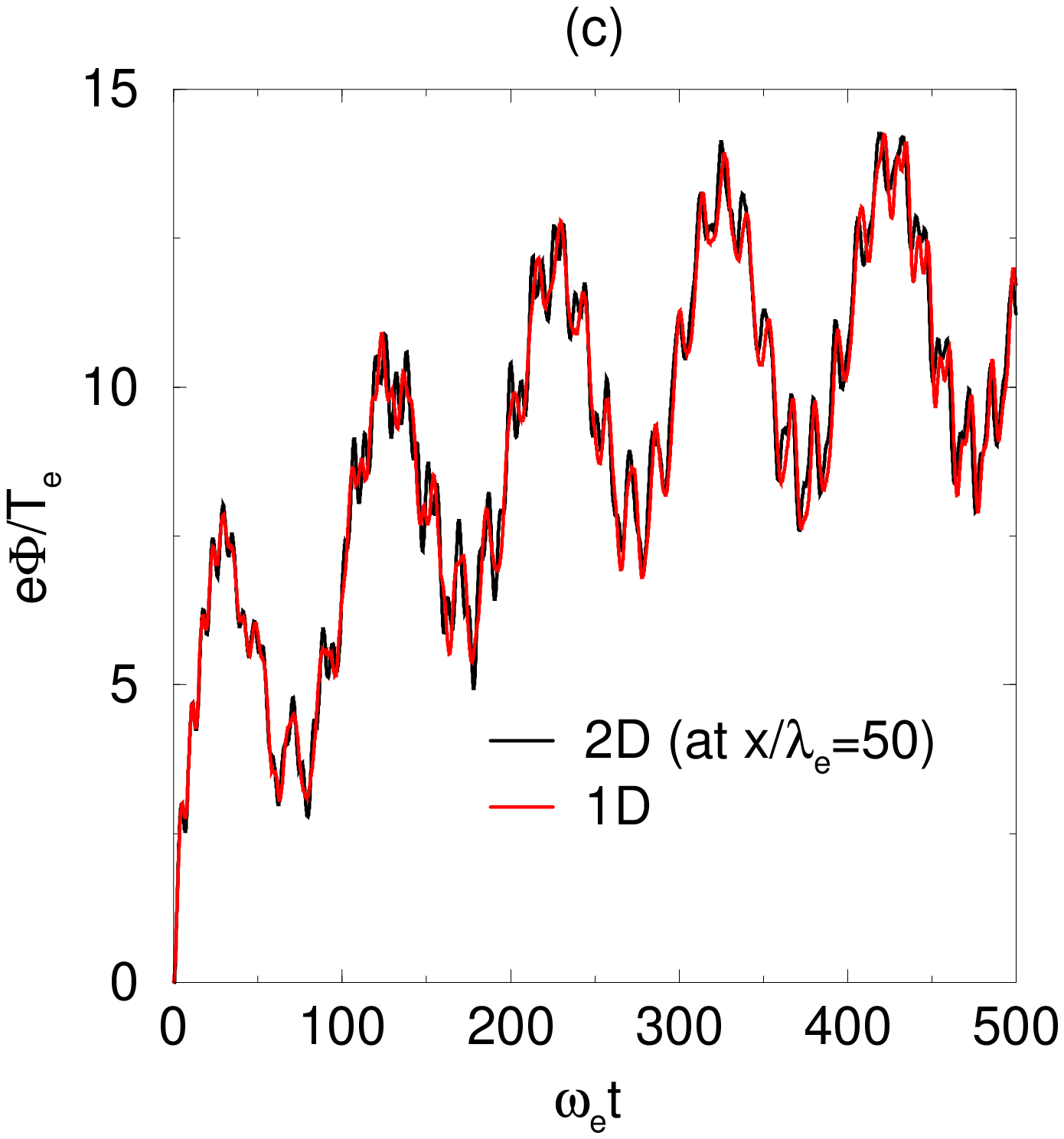}
\caption{
(a) A two dimensional potential profile at $\omega_e t = 435$.
(b) The potential profiles from two dimensional and one dimensional simulation 
compared at $\omega_e t = 385$ and $\omega_e t = 435$.
(c) Time evolution of potential values at $x / \lambda_e = 50$ obtained
from two dimensional PIC simulation (black curve) is compared with
one dimensional PIC simulation (red curve).
}
\end{figure}

Figure~3(a) compares the electron phase space plots of a two dimensional 
PIC simulation with all the $\bar{y}$ information projected onto $\bar{x}$ (black hollow dots), 
with plots from a one dimensional PIC simulation (red dots), both at $\bar{t} = 500$.
At the steady state of the RF biased discharge,
the electrons are absent in the sheath regions.
Likewise, Fig.~3(b) is from ion phase space plots at $\bar{t} = 500$, which demonstrates
ions' rapid acceleration toward the boundaries.
\begin{figure}[ht]
\centering
\includegraphics[width=2.1in]{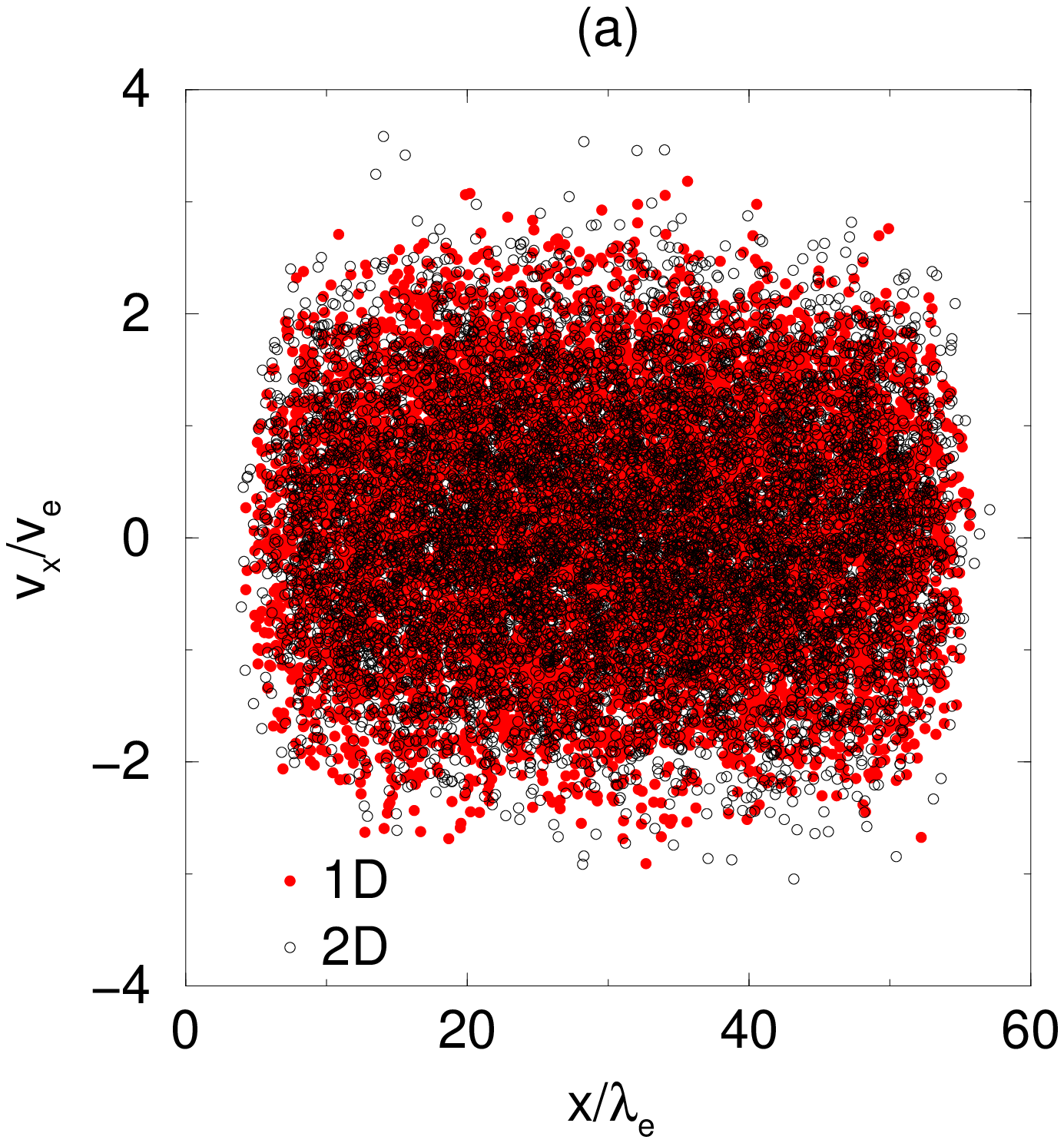}
\includegraphics[width=2.1in]{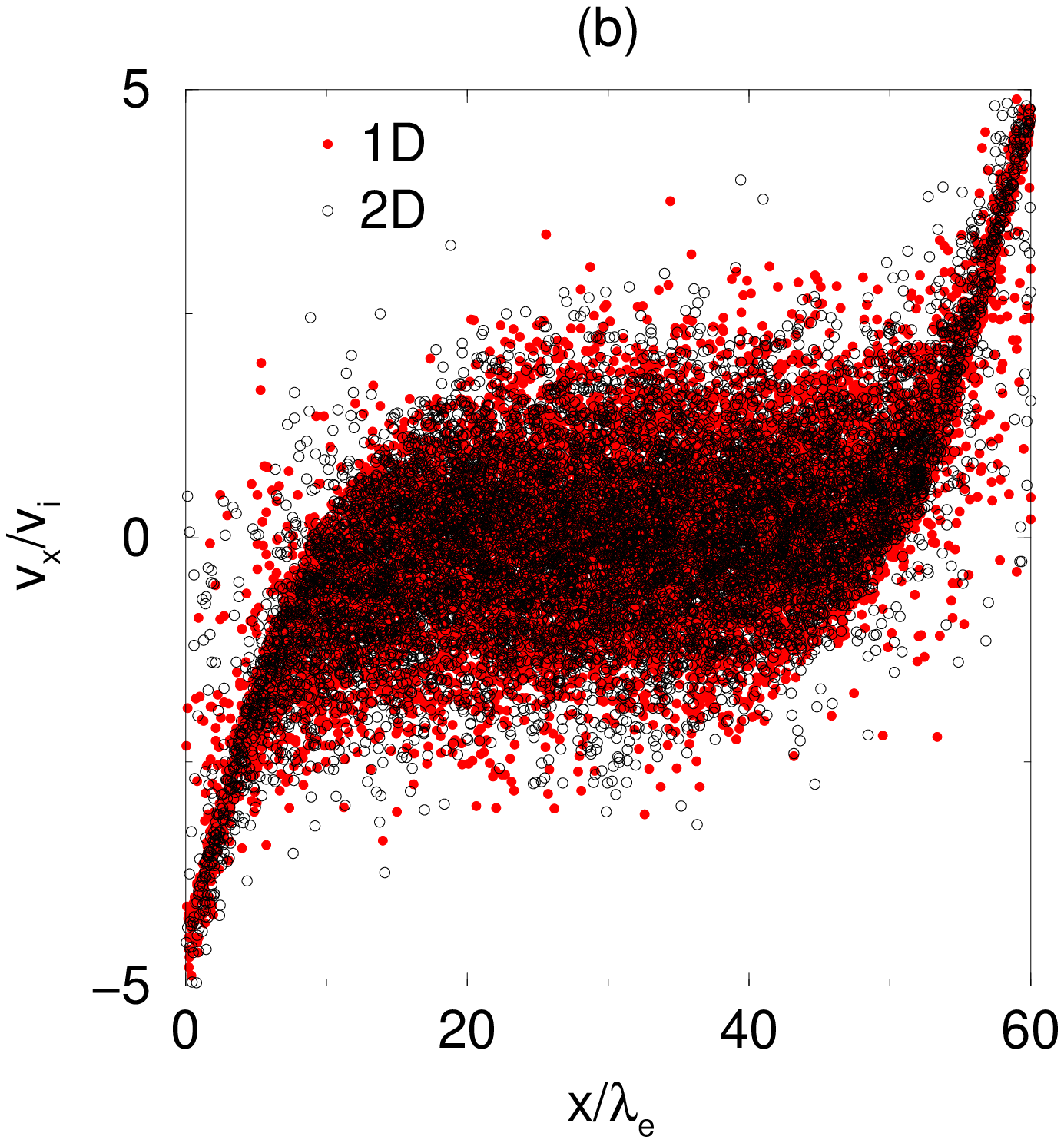}
\caption{(a) The electrons phase space plots at $\omega_e t = 500$ from 
a two dimensional PIC and a one dimensional PIC simulation.
(b) The ion phase space plots at $\omega_e t = 500$.
Black hollow dots from two dimensional and red dots from one dimensional simulation.
Eight percent of the total particles are plotted.}
\end{figure}

Interestingly, we can already
reveal the ions' velocity anisotropy for those reaching the plate on the right
(employing the results shown in Figs.2 and 3).
The two dimensional velocity distribution of ions which have reached 
the right boundary, $\bar{x}=\bar{x}_{max}$, is plotted in Fig.4. 
In the initial phase, before the sheath is formed (the hollow black dots in 
Fig.4 is from $0 \le \bar{t} \le 100$), the ion velocities are rather isotropic. 
After the sheath formation (black solid dots in Fig.4 obtained within $400 \le \bar{t} \le 500$),
when the ions are accelerated toward the plate, we see an enhanced anisotropy with the
$\bar{v}_{x}$ component being dominant. 
\begin{figure}[ht]
\centering
\includegraphics[width=2.1in]{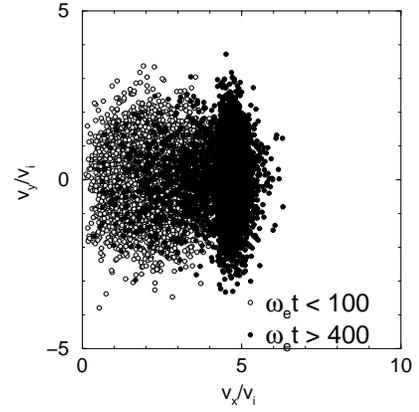}
\caption{
The ion energy spectrum which has reached and bombarded the boundary at $x/\lambda_e=60$.
An enhanced anisotropy is observed for $\omega_e t \ge 400$. Note that the velocity
is plotted in terms of normalization by the ion thermal velocity $v_i$.
}
\end{figure}

\subsection{PIC simulation with a trench geometry}
\label{s3c}
Finally, we incorporate the trench geometry. 
The size of the plasma is same as in \ref{s3b} with an additional trench length of 
$20 \lambda_e$ given in the $\bar{x}$ direction (see Figs.~5 and 6). 
The trench width in the $\bar{y}$ direction is given by $15 \lambda_e$ in Fig.~5,
and $4 \lambda_e$ in Fig.~6, respectively.
In Figs.~5 and 6, the bulk region given by $0 \le \bar{x} \le 60$ is the periodic region as in \ref{s3b}. 
Note that we recycle the ions hitting the walls as before, but the ionization is assumed 
only in the bulk region.

In Fig.5, a potential profile is developed in the $\bar{y}$ direction.
This is because the trench width is much larger than the Debye length. 
Finite potential height within the trench can be observed in Fig.5(a).
The $\bar{E}_y$ component of the electric field in the trench
causes ions to be accelerated toward the sidewalls (which is not preferred
from a plasma etching point of view). 
In Fig. 5(b), velocity distributions of the ions which have reached ({\it hit})
the trench bottom (at $\bar{x}=80$ and $5 \le \bar{y} \le 20$, black dots), 
the lower side wall (at $65 \le \bar{x} \le 80$ and $\bar{y}=5$, red dots),
and the upper side wall (at $65 \le \bar{x} \le 80$ and $\bar{y}=20$, green dots)
are plotted. The velocity data are taken during the period of $600 \le \bar{t} \le 1000$.

Accumulated numbers of particles reaching the bottom ($N_{bottom}$) and the two sidewalls 
($N_{sidewalls}$) are shown in Fig.5(c).
During the period $600 \le \bar{t} \le 1000$, 
when we have constant ion fluxes toward the trench surface 
(meaning $dN_{bottom}/dt$ and $dN_{sidewalls}/dt$ are constant), 
$2137$ ions reach the bottom and $4920$ ions reach the sidewalls
(the ratio between them are $2.31$).
\begin{figure}[ht]
\centering
\includegraphics[width=2.1in,angle=-90]{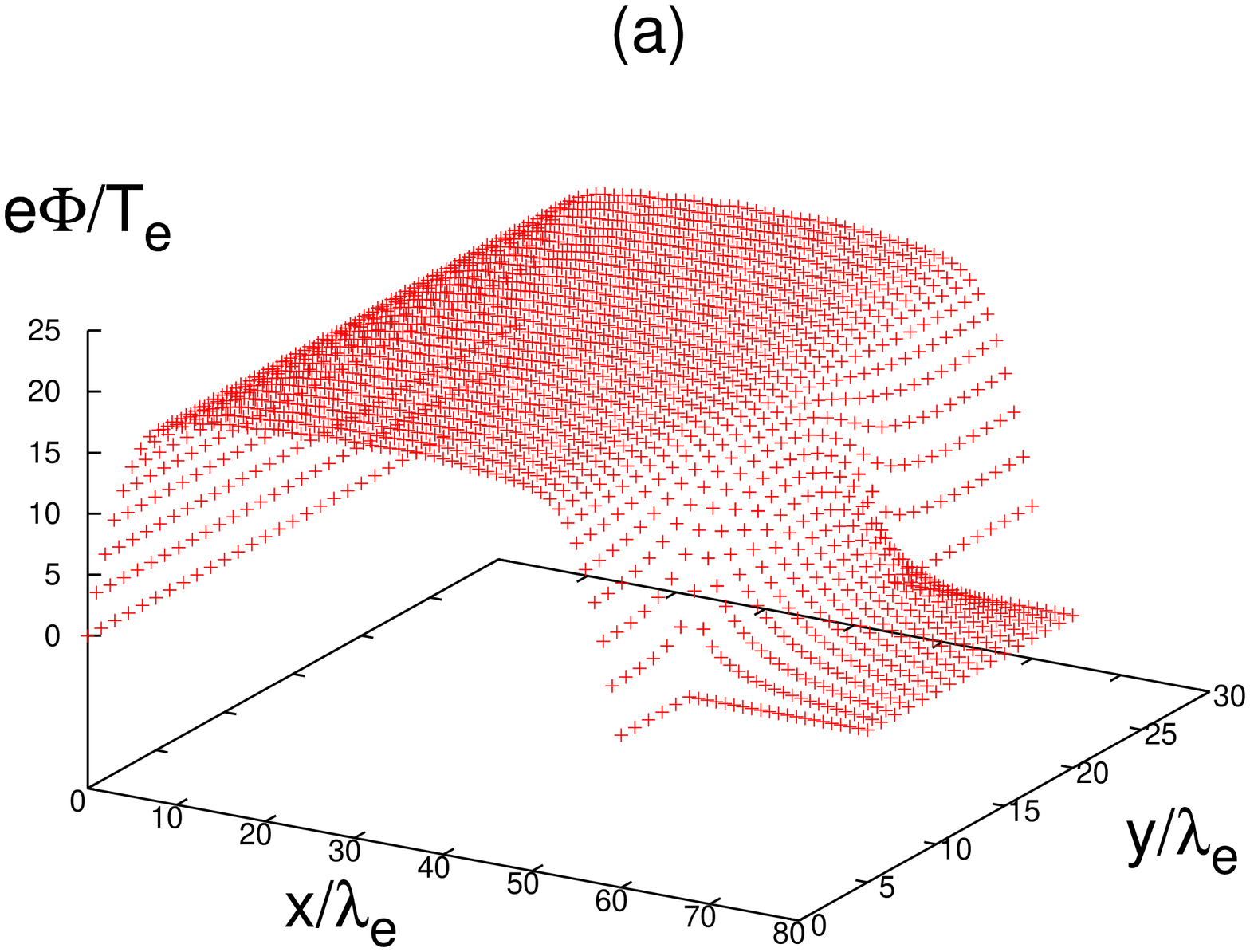}
\includegraphics[width=2.1in,angle=-00]{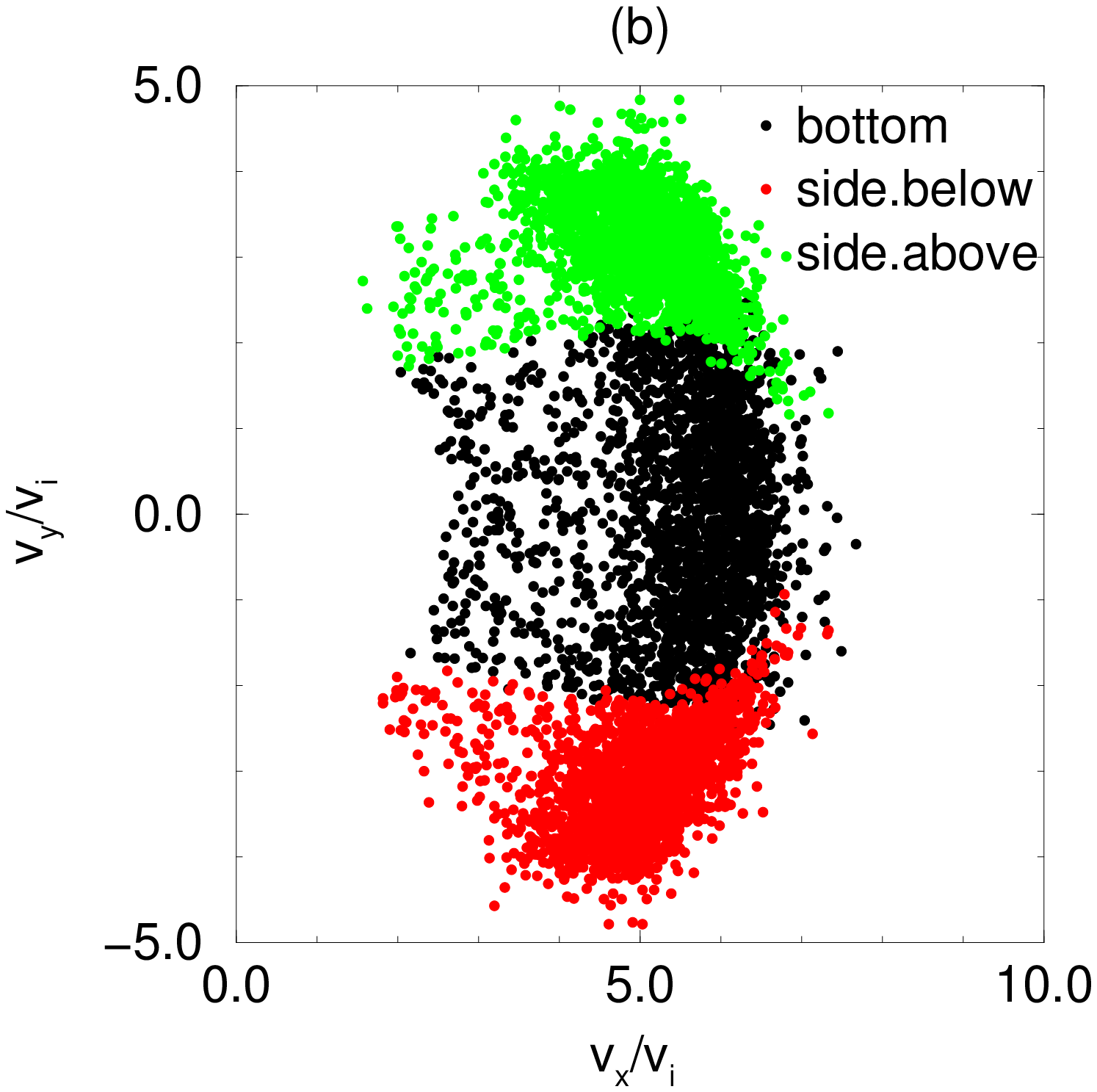}
\includegraphics[width=2.1in,angle=-00]{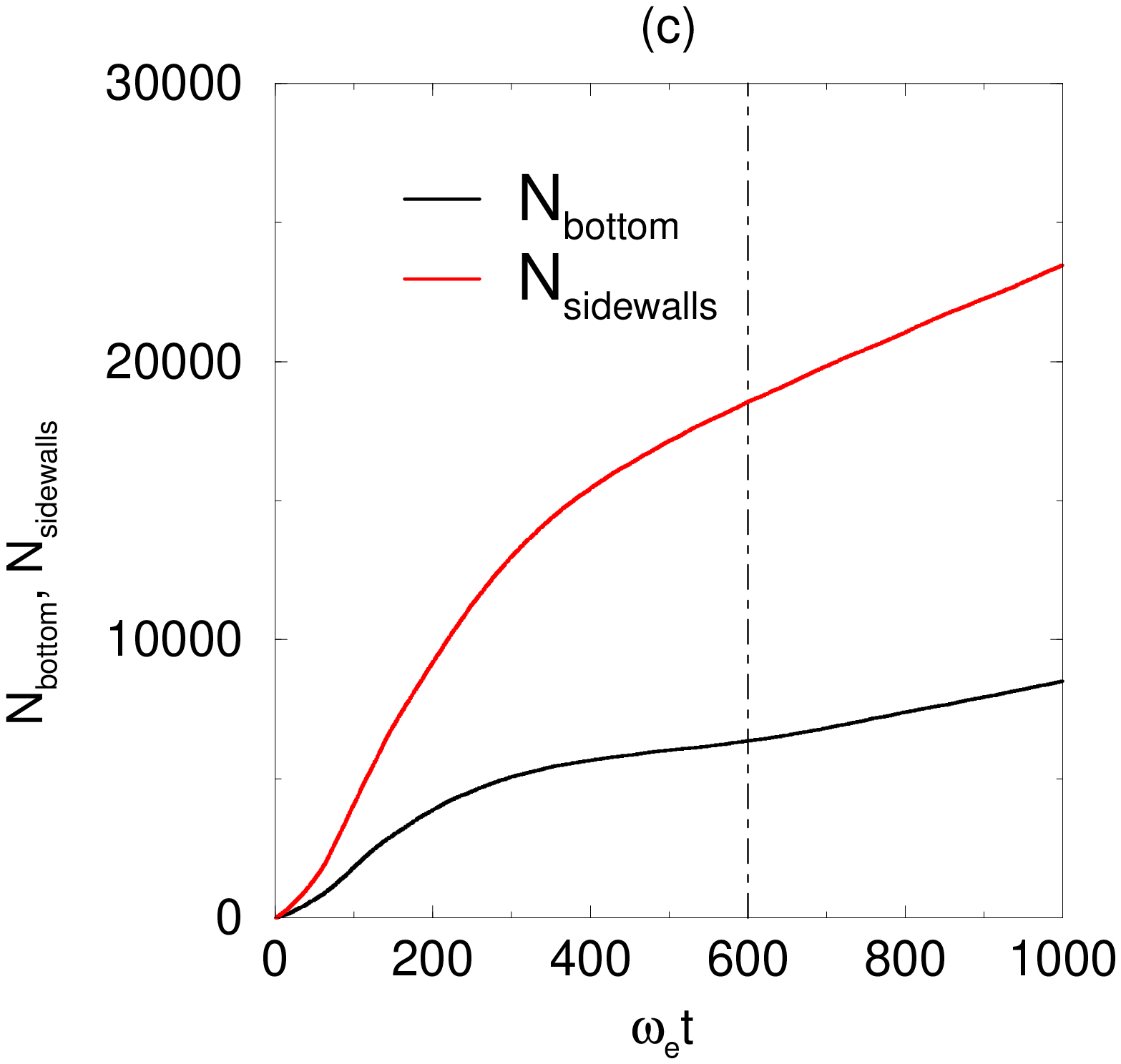}
\caption{
A two dimensional PIC simulation results for a wide trench (width of $15 \lambda_e$).
(a) A two dimensional potential profile at $\omega_e t = 1000$.
(b) The velocity distributions of the ions which have reached  
the trench bottom (black dots), 
the lower side wall (red dots),
and the upper side wall (green dots) during the period of $600 \le \omega_e t \le 1000$.
(c) Accumulated numbers of particles reaching the bottom (black curve) and the two sidewalls 
(red curve) versus time.
}
\end{figure}

A similar analysis is done for a relatively small trench width of $4 \lambda_e$ in Fig.~6. 
The trench bottom is located at $\bar{x}=80$ and $10 \le \bar{y} \le 14$.
In a thin trench region, potential profile is ignorable [see Fig.6(a)].
Ions can directly bombard the bottom of the trench.
Velocity distributions of the ions at the boundaries are given in Fig.~6(b), 
and accumulated numbers of particles are given in Fig.~6(c).
The velocity anisotropy in Fig.~6(b) is much clear compared to Fig.~5(b).
During the period $600 \le \bar{t} \le 1000$ (as in Fig.5),
$831$ ions reach the bottom and $1558$ ions reach the sidewalls
(the ratio between them are $1.85$).
By comparing Fig.5 and Fig.6, what we learn is that 
the control of the anisotropy can be related to
the competition of sheath dynamics between two perpendicular directions
which are "$\bar{x}$" (direction toward the plates) 
and "$\bar{y}$" (direction across the trench).

\begin{figure}[ht]
\centering
\includegraphics[width=2.1in,angle=-90]{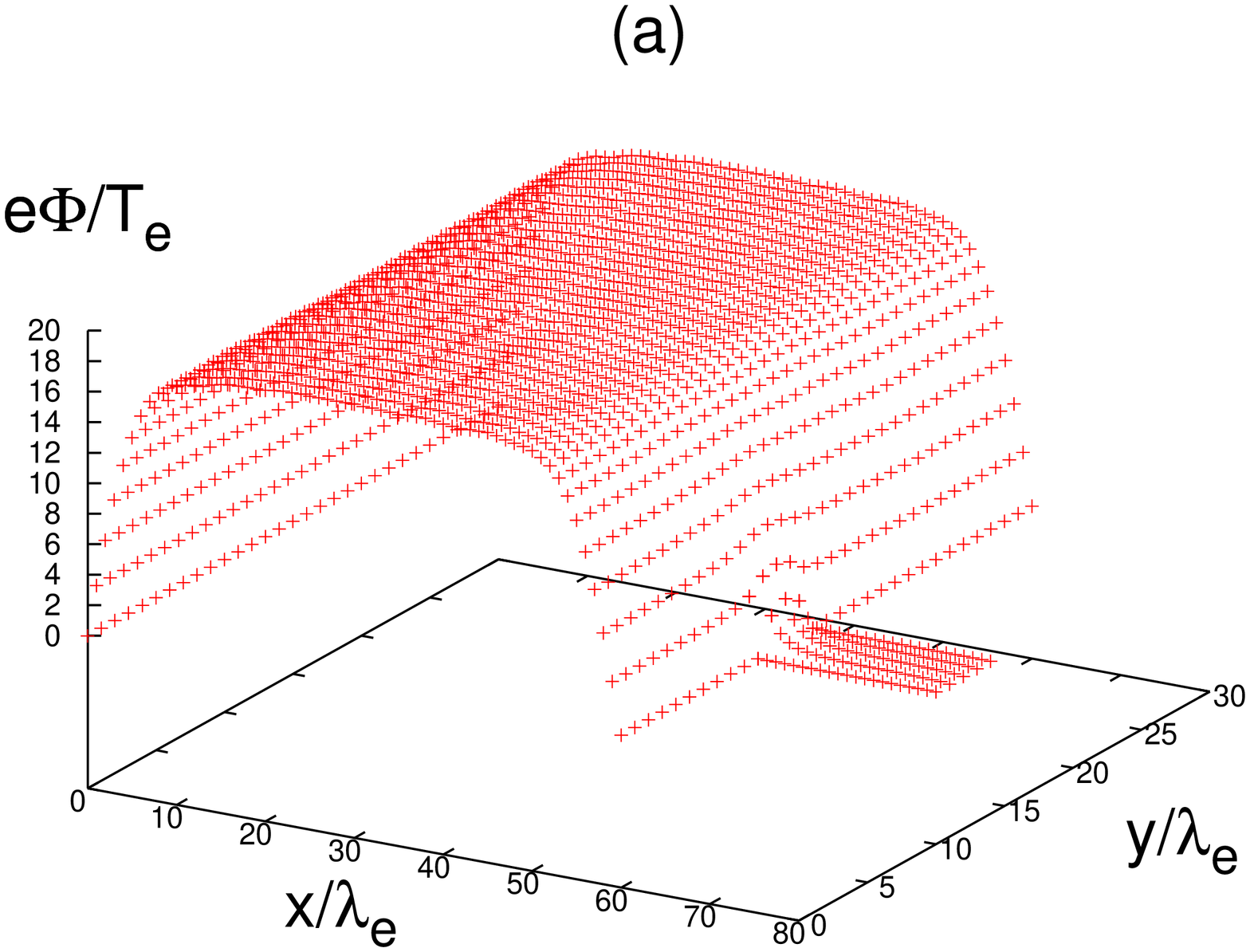}
\includegraphics[width=2.1in,angle=-00]{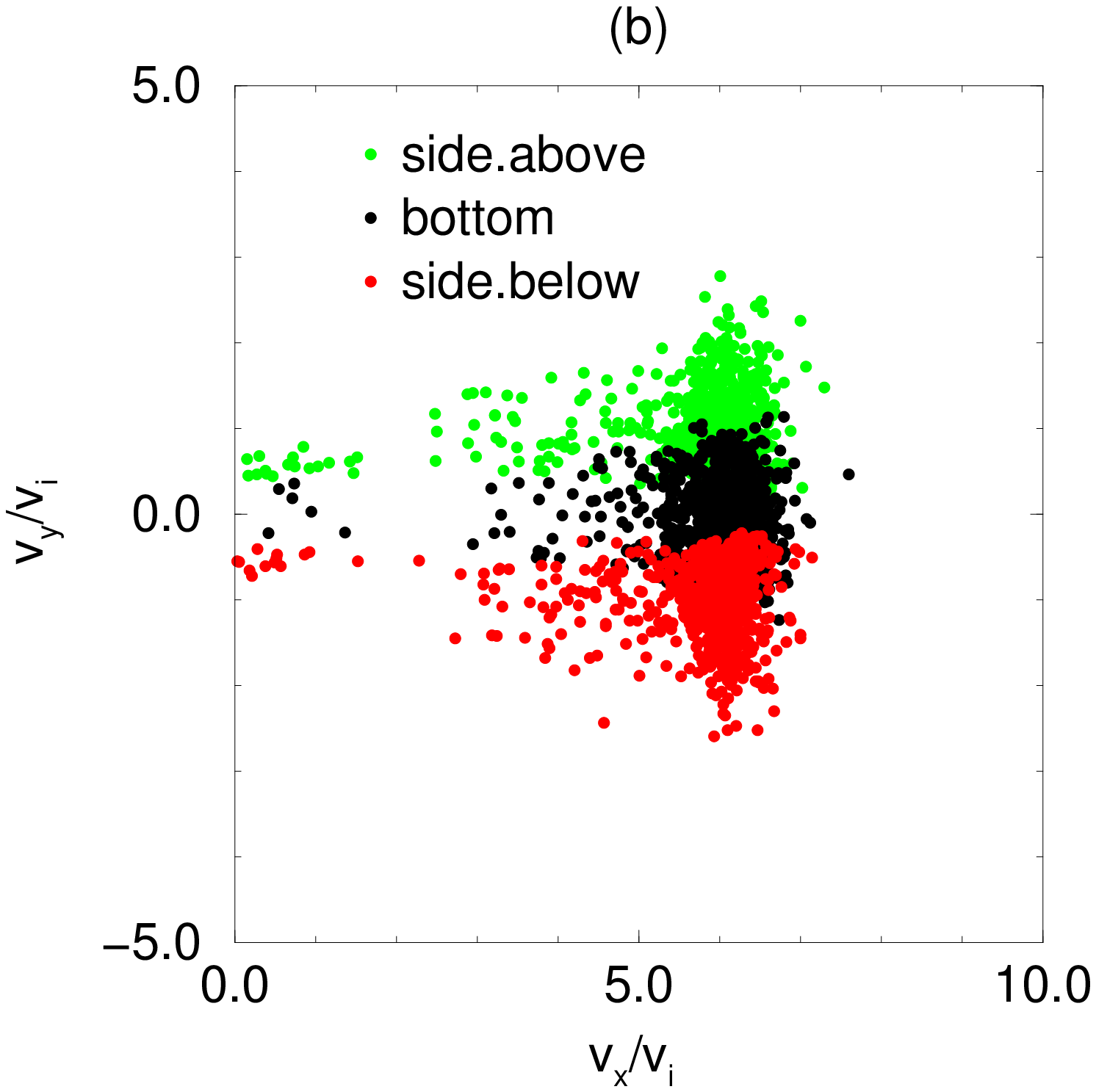}
\includegraphics[width=2.1in,angle=-00]{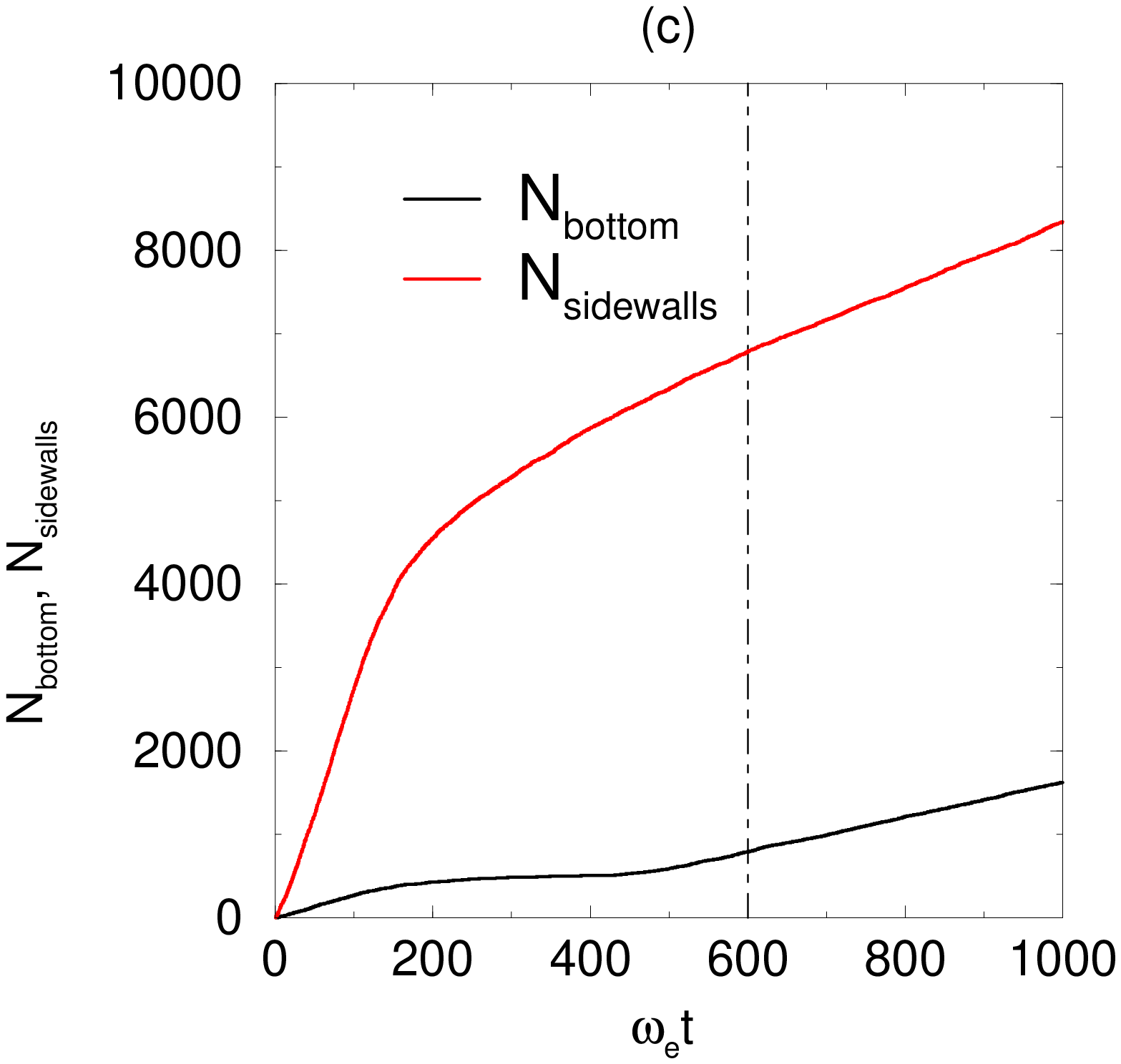}
\caption{
A two dimensional PIC simulation results for a relatively thin trench (width of $4 \lambda_e$).
(a) A two dimensional potential profile at $\omega_e t = 1000$.
(b) The velocity distributions of the ions which have reached  
the trench bottom (black dots), 
the lower side wall (red dots),
and the upper side wall (green dots) during the period of $600 \le \omega_e t\le 1000$.
(c) Accumulated numbers of particles reaching the bottom (black curve) and the two sidewalls 
(red curve) versus time.
}
\end{figure}

\subsection{Charge build up effects}
\label{s3d}

As we discussed in \ref{s1}, we are also interested in the effect of 
charge build-ups on the trench surface \cite{hit99}. 
If ions keep hitting the trench region, it is possible
that the trench bottom become positively charged and thus the potential can rise. 
The potential at the bottom of trench can be positive (relative to the trench region).
Then some of the ions are repelled which  in turn can gives rise to the sidewall erosion.
Here, on top of the analysis in Fig.5, we provide the plasma with
additional stationary ion charge density at the trench bottom.
We provide an extra test ion density of $\bar{n}_i = 5.0$ 
at $\bar{x}=79$ and $5 \le \bar{y} \le 20$
(we do not time advance this latter portion of the ion density).
 
Figure 7(a) shows potential profile at $\bar{t} = 1000$ which demonstrates 
significant potential variation in the trench region compared to the previous two cases. 
The potential gradient cause electric field to decelerate ions 
which was originally approaching the trench bottom.
As in Fig.~5, velocity distributions are given in Fig.~7(b),
which suggest loss of anisotropy [see Fig.~5(b) as well]. 
The accumulated numbers of particles are given in Fig.~7(c):
the solid curves are the ones with charged build-ups and the dashed curves
are the ones from Fig.~5(c) for comparison.
The test simulation result shows much less ions are reaching
the bottom but instead more ions are reaching the sidewalls.
During the period of $700 \le \bar{t} \le 1000$,
$1094$ ions reach the bottom and $3820$ ions reach the sidewalls
(the ratio between them are $3.49$. 
As a reminder, smaller ratios suggest better anisotropy).

\begin{figure}[ht]
\centering
\includegraphics[width=2.1in,angle=-90]{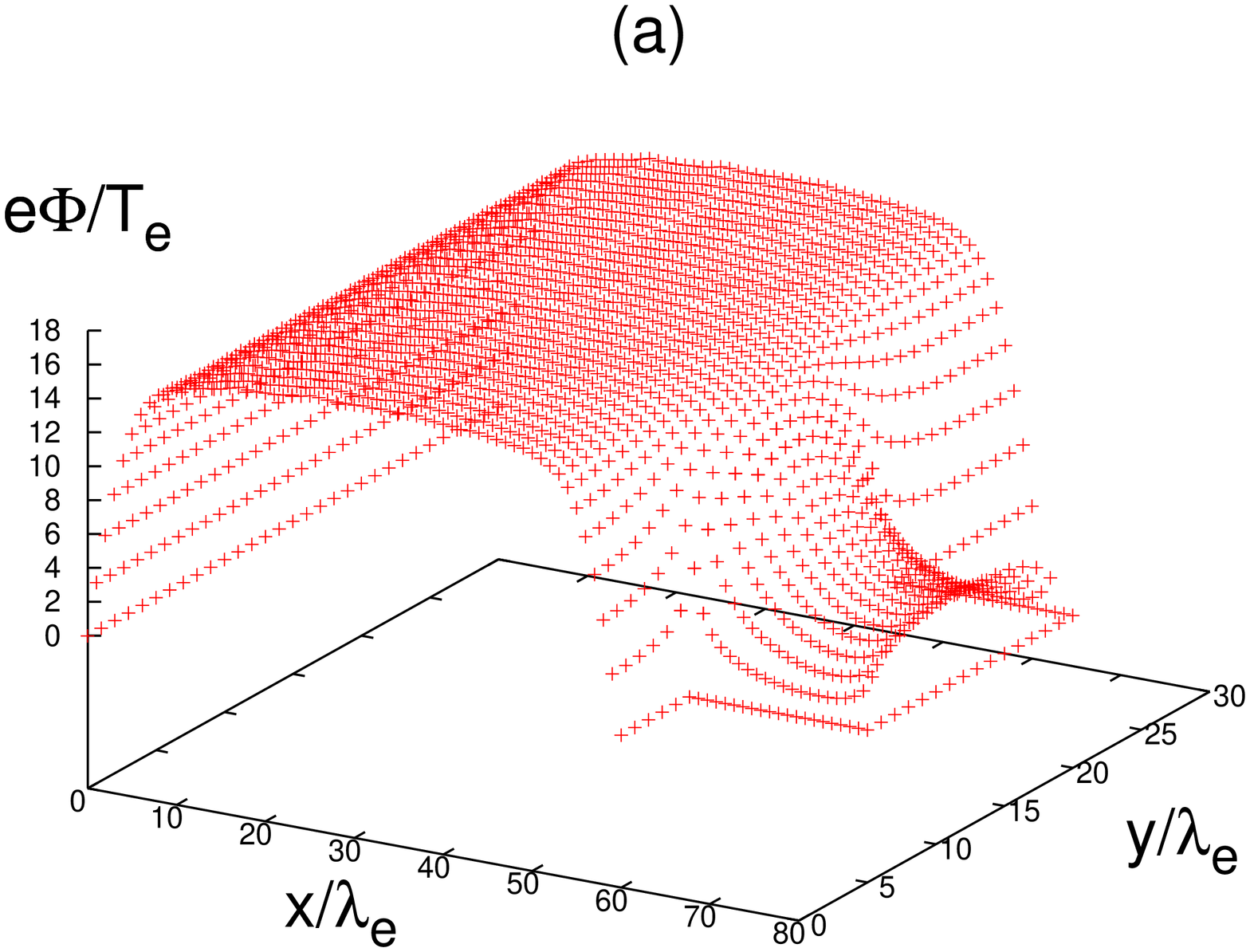}
\includegraphics[width=2.1in,angle=-00]{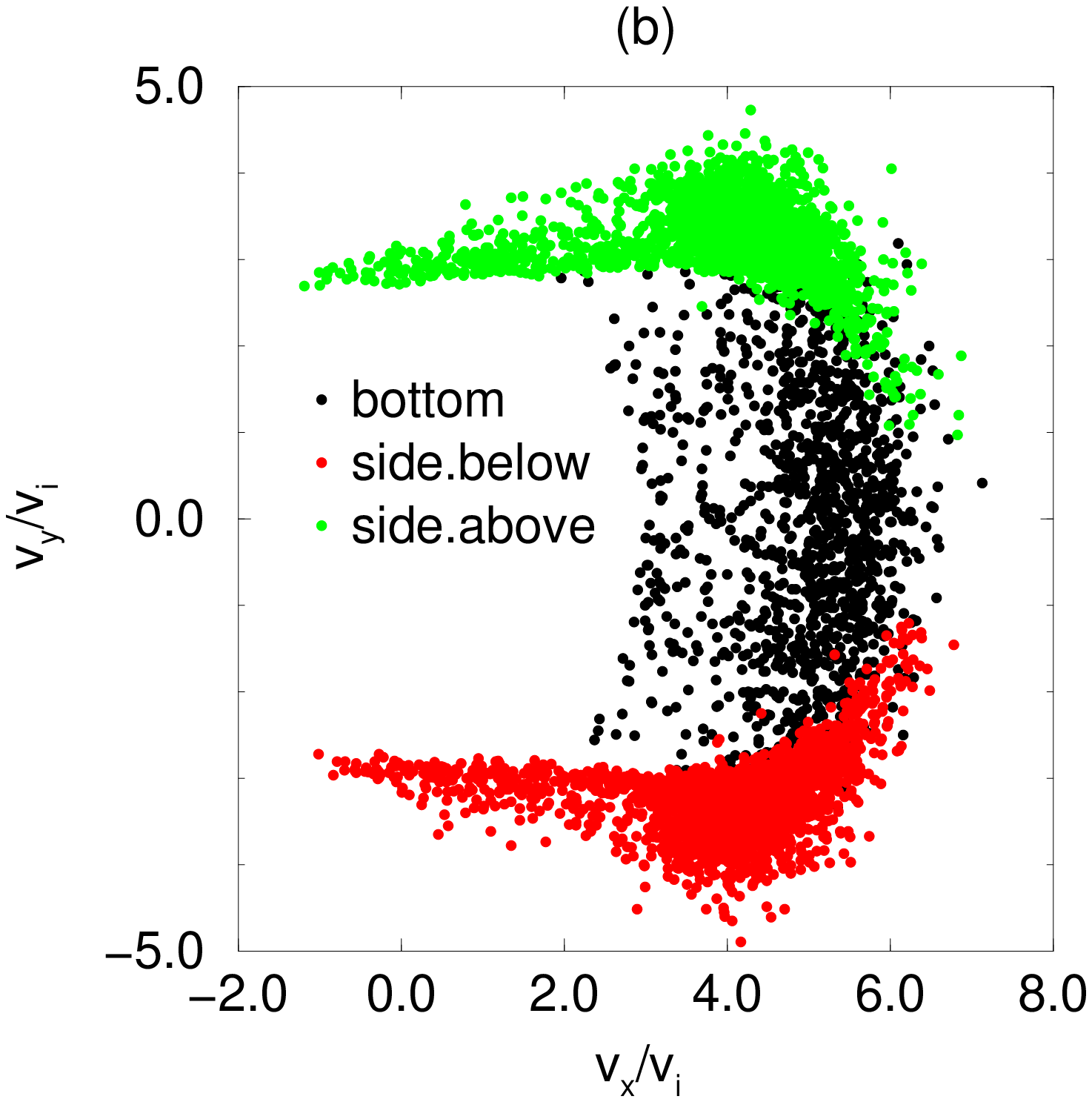}
\includegraphics[width=2.1in,angle=-00]{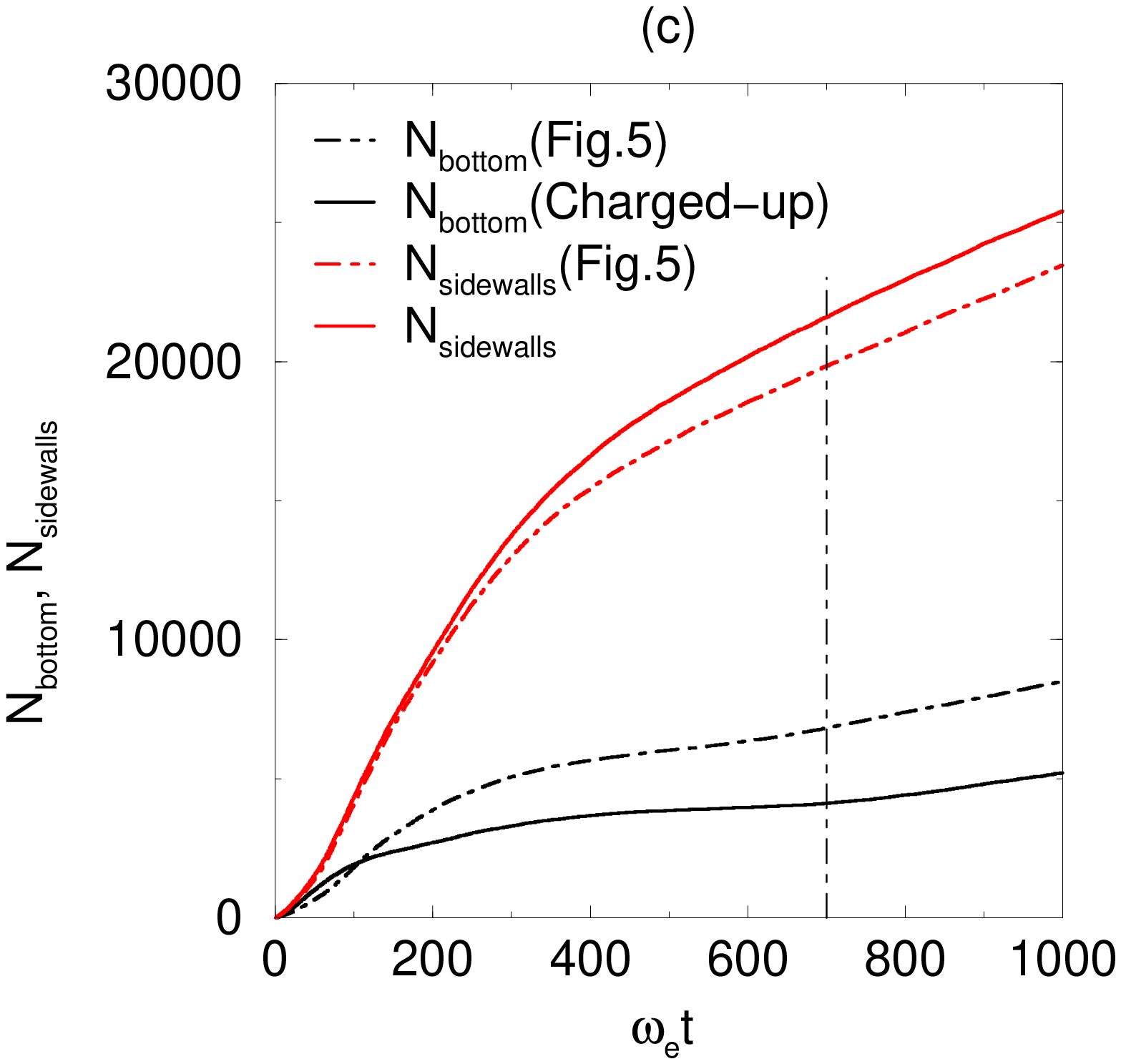}
\caption{
A two dimensional simulation results with a charge build up at the trench
bottom. The rest of the setting are the same as in Fig.5.
(a) A two dimensional potential profile at $\omega_e t = 1000$.
(b) The velocity distributions of the ions which have reached  
the trench bottom (black dots), 
the lower side wall (red dots),
and the upper side wall (green dots) during the period of $700 \le \omega_e t \le 1000$.
(c) Accumulated numbers of particles reaching the bottom (black curve) and the two sidewalls 
(red curve) versus time.
}
\end{figure}

\section{Summary and discussions}
\label{s4}
In this work, we have built a two-dimensional electrostatic PIC model 
to examine anisotropy of ion bombardment in a trench geometry \cite{hit99}.
We made our simulation results gradually closer to realistic plasma discharge processes. 
We have started from a two dimensional square shape with Dirichlet boundary conditions, 
then advanced to a square geometry with a periodic boundary condition, and finally
added the trench geometry.
The simulation results indicate ions can drift
toward the sidewalls when the potential profile is developed in the trench region when the
width is larger than a few Debye length. 
To the contrary, anisotropy can be achieved in a relatively small trench size.
Furthermore, positive charging up on the trench bottom has been examined. The
numerical simulation result indicates that if we have positive potential due to ions’ accumulation on the
trench bottom, we can lose anisotropy.

To focus on the basic geometrical effects of the trench in the first place,
the trench size we dealt in this work are larger than conventional ones 
(while our bulk plasma size is comparable to the industrial processing plasmas).
In the future, we would like to extend our two dimensional PIC simulation
to a sub-micron trench with a further speed-up of the
numerical computation, the field solver in particular \cite{nis06}.
On the other hand, another path is to develop
an N-body simulation method (as in gravitational N-body problems \cite{mak98}) 
by calculating individual Coulomb force between charged particles.

Generally, secondary electron emission is not a preferable ingredient for etching.
It causes the sheath instability and enhance power loss \cite{sur91,cam12a,cam12b,cam16}. 
However, in the trench region, it is possible that the
secondary electron emissions reduce the plasma potential. 
Ions affected by sidewalls can be reduced.  
Inclusion of the secondary electron emission effects into
our two dimensional model through the ion bombardment \cite{sur91}
and the electron bombardment \cite{cam12a,hua15} is yet our another
near future task. 

\section*{Acknowledgment}
The authors would like to thank 
Dr. M.~R.~Smith and Dr. F.~C.~N.~Hong of National Cheng Kung University, 
and Dr. I.~Kaganovich of Princeton University for useful discussions.
This work is supported by Taiwan MOST 103-2112-M-006-007 and MOST 104-2112-M-006-019.

\vspace{-1.5cm}
\begin{IEEEbiographynophoto}{Tai-Lu Lin}
has completed M.S. degree at the Institute of Space and Plasma Sciences (ISAPS), National Cheng
Kung University (NCKU), Tainan, Taiwan in February 2016.
His thesis has focused on plasma kinetic theory in industrial plasmas. 
He is now with Chunghwa Picture Tubes, LTD. in Taoyuan, Taiwan.

{\bf Yasutaro Nishimura}
received the Ph.D. degree from the University of
Wisconsin-Madison, Madison, in 1998.
He has worked with the Max-Planck-Institut f\"{u}r Plasmaphysik, Garching,
Germany, and the University of California. He is currently an Associate
Professor with the ISAPS, NCKU, Tainan, Taiwan, where he focuses on kinetic instabilities in 
high temperature magnetized plasmas.
\end{IEEEbiographynophoto}

\end{document}